\documentclass[preprint,superscriptaddress,amsmath, nofootinbib]{revtex4-1}
\usepackage{graphicx}
\usepackage{latexsym}
\usepackage{slashed}
\usepackage{amsmath}
\usepackage{bm}
\usepackage{color}
\usepackage{xcolor}
\usepackage{booktabs}
\usepackage[colorlinks,citecolor=blue,urlcolor=blue,linkcolor=blue]{hyperref}
\usepackage{diagbox}
\usepackage{textcomp}

\definecolor{nicered}{rgb}{0.7,0.1,0.1}
\definecolor{nicegreen}{rgb}{0.1,0.5,0.1}
\hypersetup{colorlinks,citecolor= nicegreen,linkcolor= nicered}

\usepackage{dcolumn}
\usepackage{bm}
\usepackage{slashed}

\begin{document}

\title{Probing Inelastic Dark Matter at the LHC, FASER and STCF}

\author{Chih-Ting Lu}
\email{ctlu@njnu.edu.cn}
\affiliation{Department of Physics and Institute of Theoretical Physics, Nanjing Normal University, Nanjing, 210023, China}
\affiliation{CAS Key Laboratory of Theoretical Physics, Institute of Theoretical Physics, Chinese Academy of Sciences, Beijing 100190, P. R. China}

\author{Jianfeng Tu}
\email{tujf@nnu.edu.cn}
\affiliation{Department of Physics and Institute of Theoretical Physics, Nanjing Normal University, Nanjing, 210023, China}

\author{Lei Wu}
\email{leiwu@njnu.edu.cn}
\affiliation{Department of Physics and Institute of Theoretical Physics, Nanjing Normal University, Nanjing, 210023, China}

\date{\today}

\begin{abstract}
In this work, we explore the potential of probing the inelastic dark matter (DM) model with an extra $U(1)_D$ gauge symmetry at the Large Hadron Collider, ForwArd Search ExpeRiment and Super Tau Charm Factory. To saturate the observed DM relic density, the mass splitting between two light dark states has to be small enough, and thus leads to some distinctive signatures at these colliders. By searching for the long-lived particle, the displaced muon-jets, the soft leptons, and the mono-photon events, we find that the inelastic DM mass in the range of 1 MeV to 210 GeV could be tested.
\end{abstract}
\pacs{Valid PACS appear here}
\maketitle

\tableofcontents

\section{Introduction}

{Despite the strong astrophysical and cosmological evidence to support the existence of dark matter (DM)~\cite{Trimble:1987ee,Barack:2018yly}, its nature still remains a mystery. It is widely believed that Standard Model (SM) particles may interact with DM through forces other than gravity~\cite{Pospelov:2007mp}. The Weakly Interacting Massive Particle (WIMP) is one of the most popular dark matter candidates~\cite{Feng:2010gw,Bauer:2017qwy}. However, null results of searching for WIMPs have imposed stringent constraints on its properties in the mass range of GeV to TeV~\cite{Schumann:2019eaa,Kahlhoefer:2017dnp}. For instance, the spin-independent (SI) DM-nucleon scattering cross section is limited to $6.5\times10^{-48}$ $\text{cm}^2$ at the DM mass of $30$ GeV~\cite{LZ:2022ufs,XENON:2023sxq}. 
This motivates the recent studies of the light DM models that can avoid the conventional constraints. To explore the sub-GeV DM, various new proposals and experiments have been proposed~\cite{Knapen:2017xzo,Lin:2022hnt}.
}

However, in the thermal freeze-out scenario, the sub-GeV DM models usually suffer from various astrophysical and cosmological constraints. For example, the s-wave annihilation of light DM is not favored by the cosmic microwave background (CMB), which requires the DM mass should be heavier than about $10$ GeV~\cite{Lin:2011gj}. Additionally, the DM particles with the mass less than $1$ MeV is tightly constrained by the Big Bang Nucleosynthesis (BBN)~\cite{Nollett:2014lwa}. Nevertheless, there exist some exceptions that can evade these bounds, such as the DM models with the p-wave annihilation~\cite{Matsumoto:2018acr}, the inelastic DM models~\cite{Tucker-Smith:2001myb}, the asymmetric DM models~\cite{Zurek:2013wia}, and the freeze-in mechanism DM models~\cite{Dvorkin:2020xga}. Among them, the inelastic DM models that were motivated by the explanation of the DAMA excess~\cite{Tucker-Smith:2001myb} have recently gained considerable attention~\cite{Baek:2014kna,Izaguirre:2015zva,DEramo:2016gqz,Izaguirre:2017bqb,Berlin:2018jbm,Mohlabeng:2019vrz,Tsai:2019buq,Okada:2019sbb,Ko:2019wxq,Duerr:2019dmv,Duerr:2020muu,Ema:2020fit,Kang:2021oes,Bell:2021zkr,Batell:2021ooj,Bell:2021xff,Feng:2021hyz,Guo:2021vpb,Li:2021rzt,Filimonova:2022pkj,Bertuzzo:2022ozu,Gu:2022vgb,Li:2022acp,Mongillo:2023hbs,Heeba:2023bik}. If only the lighter DM state is present in the current Universe, the up-scattering in DM-nucleon interactions becomes insensitive to direct detection\footnote{Note people can consider cosmic ray-boosted inelastic DM for the up-scattering in DM-nucleon interactions as shown in Ref.~\cite{Bell:2021xff,Feng:2021hyz}}, and the primary elastic DM-nucleon scattering occurs at the one-loop level~\cite{Izaguirre:2015zva}. Consequently, these models can evade the limits from direct detection~\cite{CarrilloGonzalez:2021lxm}.

In this paper, we investigate the prospect of probing the inelastic DM model with an additional $U(1)_D$ gauge symmetry at colliders. The mass splitting of two dark states is induced by the interaction between the dark Higgs field and the DM sector, and the transition between two dark states is mediated by the new $U(1)_D$ gauge boson. The collider signatures of inelastic DM at accelerators strongly depend on two key parameters: the ground state DM mass ($M_{\chi_1}$) and the mass splitting between the excited and ground DM states ($\Delta_{\chi}\equiv M_{\chi_2}-M_{\chi_1}$). These two parameters are also associated with the lifetime of the excited DM state. 
For $M_{\chi_1}\lesssim 5$ GeV and $\Delta_{\chi} < 0.5 M_{\chi_1}$, the fixed target experiments~\cite{A1:2011yso,PhysRevLett.107.191804,PhysRevLett.112.221802} and low energy $e^+ e^-$ colliders such as BaBar~\cite{Filippi:2019lfq}, Belle II~\cite{Belle-II:2022yaw}, BESIII~\cite{BESIII:2022oww} and Super Tau Charm Factory (STCF)~\cite{Epifanov:2020elk} offer powerful avenues for searching for inelastic DM. On the other hand, for $M_{\chi_1}\lesssim 15$ GeV and $\Delta_{\chi}\lesssim 0.1 M_{\chi_1}$, LLP experiments like ForwArd Search ExpeRiment (FASER)~\cite{FASER:2018bac,FASER:2019aik}, MAssive Timing Hodoscope for Ultra-Stable neutral-pArticles (MATHUSLA)~\cite{MATHUSLA:2022sze}, SeaQuest~\cite{Liu:2023buo}, the COmpact Detector for EXotics at LHCb (CODEX-b)~\cite{Aielli:2022awh} and A Laboratory for Long-Lived eXotics (AL3X)~\cite{Dercks:2018wum} can explore the remaining parameter space.
In the intermediate DM mass range of $2$ GeV $\lesssim M_{\chi_1} \lesssim 200$ GeV with $\Delta_{\chi} \lesssim 0.2 M_{\chi_1}$, the Large Hadron Collider (LHC) remains the primary machine for probing inelastic DM. 

The structure of this paper is expanding as follows. In Sec.~\ref{sec:model}, we recapitulate the inelastic DM model with an $U(1)_D$ gauge symmetry. In Sec.~\ref{sec:signature}, we then study the signatures of inelastic DM at the FASER, LHC, and STCF. Finally, we summarize our findings in Sec.~\ref{sec:conclusion}.

\section{Inelastic dark matter model} 
\label{sec:model}

{ 
In this section, we briefly review the inelastic DM models with an $U(1)_D$ gauge symmetry and focus on the fermionic DM candidates\footnote{The scalar inelastic DM models can be found in Ref.~\cite{Baek:2014kna, Izaguirre:2015zva, Okada:2019sbb, Kang:2021oes, Li:2021rzt, Bertuzzo:2022ozu}. Since we will study the on-shell $Z'$ productions and $Z'$ mainly decays to DM states, the predictions in our analysis can be applied to scalar inelastic DM models as well.}. In addition to the SM particles, a singlet complex scalar field $\Phi$ as well as a Dirac fermion field $\chi$ are involved. We assign the $U(1)_D$ charges for $\Phi$ and $\chi$ as $Q(\Phi) = +2$ and $Q(\chi) = +1$, respectively. All SM particles are neutral under the $U(1)_D$ symmetry and cannot be directly coupled to the dark sector. The relevant gauge invariant and renormalizable Lagrangian for this model can be written as 
\begin{equation}
\begin{split}
    \mathcal{L} =& \mathcal{L}_{SM}  -\frac{1}{4}X_{\mu\nu}X^{\mu\nu}-\frac{1}{2}\sin{\epsilon}X_{\mu\nu}B^{\mu\nu}+\mathcal{D}^\mu{\Phi^\dag}\mathcal{D}_\mu{\Phi} \\
    &- \mu^2_{\Phi}\Phi^\dag{\Phi}+\lambda_{\Phi}(\Phi^\dag{\Phi})^2-\lambda_{\mathcal{H}\Phi}\mathcal{H}^\dag\mathcal{H}\Phi^\dag{\Phi} \\
    &- \overline{\chi}(i\mathcal{D} \!\!\!/ - M_\chi)\chi - (\frac{\xi}{2}\Phi^{\dag}\overline{\chi^c}\chi + H.c.),
    \label{Lf}
\end{split}
\end{equation}
where $X_{\mu\nu}$ and $B_{\mu\nu}$ are field strength tensors of $U(1)_D$ and $U(1)_Y$ gauge fields, respectively. $\epsilon$ is the kinematic mixing angle between $X_{\mu\nu}$ and $B_{\mu\nu}$, $\mu_{\Phi}$ is the parameter with the same dimension as mass, and $\lambda_{\Phi}$, $\lambda_{\mathcal{H}\Phi}$ are dimensionless parameters and $\xi$ is assumed to be a positive, real and dimensionless parameter. $\mathcal{H}$ is the SM-like scalar doublet field and we expand $\mathcal{H}$ and $\Phi$ in the unitary gauge to the following form, 
\begin{align}
    {\mathcal{H}}\ = \ \frac{1}{\sqrt{2}} \begin{pmatrix}
    {0} \\
    v+h
    \end{pmatrix},\quad  \Phi=\frac{1}{\sqrt{2}}(v_X+h_X), 
\end{align}
where $v$ and $v_X$ are vacuum expectation values of $\mathcal{H}$ and $\Phi$, respectively.
The $U(1)_D$ is broken spontaneously by $\langle \Phi \rangle$ = ${v_{X}}/{\sqrt{2}}$, and electroweak symmetry is broken spontaneously as usual by $\langle \mathcal{H} \rangle$ = $(0,{v}/{\sqrt{2}})$.

We then diagonalize the $U(1)$ gauge kinetic term in Eq.~(\ref{Lf}) by redefining the gauge fields via the following transformation~\cite{Wells:2008xg,Berlin:2018jbm,Filimonova:2022pkj} :
\begin{equation}
\left(
\begin{array}{cc}
    B^{\mu}  \\
    X^{\mu}
\end{array}
\right)
=
\left(
\begin{array}{cc}
     1 & \eta \\
     0 & \eta/\epsilon
\end{array}
\right)
\left(
\begin{array}{cc}
     B^{\mu}_p  \\
     X^{\mu}_p
\end{array}
\right).
\end{equation}
The covariant derivative is given by $\mathcal{D}_{\mu} = \partial_{\mu} + i(g_{D}Q_{X}+ g_{1}\eta Q_{Y})X_{\mu}+ig_{1}Q_{Y}B_{\mu}+ig_{2}T^{3}W^{3}_{\mu}$,
where $W^{3}_{\mu}$, $B_{\mu}$, and $X_{\mu}$ correspond to the gauge potentials associated with the gauge groups $SU(2)_{L}$, $U(1)_{Y}$, and $U(1)_{X}$, respectively. The gauge couplings are denoted as $g_{2}$, $g_{1}$, and $g_{D}$. The gauge fields before mixing are represented by $B^{\mu}_p$ and $X^{\mu}_p$. Additionally, we define $\eta \equiv{} \epsilon/\sqrt{1-\epsilon^2}$, and $Q_X$ represents the $U(1)_D$ charge of either $\Phi$ or $\chi$.

After performing a GL$(2, R)$ rotation to diagonalize the kinetic terms, followed by an $\textit{O}(3)$ rotation to diagonalize the $3\times 3$ neutral gauge boson mass matrix, the mass eigenstates can be expressed through the corresponding transformation as~\cite{Filimonova:2022pkj} :
\begin{equation}
\left(
\begin{array}{cc}
    B_{p}^{\mu}  \\
    W^3  \\
    X_{p}^{\mu}
\end{array}
\right)
=
\left(
\begin{array}{ccc}
     c_W & -s_W c_X & s_W s_X \\
     s_W & c_W c_X & -c_W s_X \\
     0 & s_X & c_X
\end{array}
\right)
\left(
\begin{array}{ccc}
     A  \\
     Z  \\
     Z^\prime
\end{array}
\right),
\end{equation}
where the $s_W $ and $c_W$ are the sine and cosine values of the weinberg angle, and the new gauge mixing angle can be written as 
\begin{equation}
    \theta_X = \frac{1}{2}\arctan(\frac{-2s_W\eta}{1-s_W^2\eta^2-\Delta_Z}),
\end{equation}
where $\Delta_Z = M_X^2/M_{Z_{0}}^2$, $M^2_{X} = g^2_{D}Q^2_{X}v_{X}^2 $ and $M^2_{Z_{0}} = (g^2_{1}+g^2_{2}) v^2/4$, {where} $ M_X$ and $M_Z$ are the masses of two $U(1)$ gauge bosons before mixing. Finally, the photon becomes massless, and two heavier gauge boson mass eigenvalues are
\begin{equation}
    M_{Z,Z'} = \frac{M_{Z_{0}}^2}{2}[(1+s_W^2\eta^2+\Delta_Z)\pm\sqrt{(1-s_W^2\eta^2+\Delta_Z)^2+4s_W^2\eta^2}], 
    \label{U1mass}
\end{equation} 
which valid for $\Delta_Z < 1-s_W^2\eta^2$. Considering the assumption $\epsilon \ll 1$, we find $M_{Z'} \approx M_X$ from Eq.~(\ref{U1mass}) and the interactions of $Z'$ and SM fermions for the linear order approximation in $\epsilon$ can be written as  
\begin{equation}
    \mathcal{L}_{Z'\overline{f}f} = -\epsilon{e}{c_W}{Q_f}\overline{f}\gamma^{\mu}f Z'_\mu,
    \label{Lzff}
\end{equation}
where $Q_f$ is the electric charge of SM fermions.

The Dirac fermion field can be further decomposed into two Majorana fermion fields $\chi_1$, $\chi_2$ as 
\begin{equation}
\begin{split} 
    \chi &= \frac{1}{\sqrt{2}}(\chi_2+i\chi_1), \\
    \chi_2 &= \chi^c_2,\quad \chi_1 = \chi^c_1.
\end{split}
\end{equation}
After the breaking of $U(1)_D$ gauge symmetry, the DM parts of Eq.~(\ref{Lf}) can be expanded as 
\begin{equation}
\begin{split}
    \mathcal{L}_{\chi} =& \frac{1}{2}\sum_{n=1,2}\overline{\chi_n}(i\partial \!\!\!/ -M_{\chi})\chi_n -i\frac{g_D}{2}{\left(\overline{\chi_2} {X \!\!\!\!/}\chi_1 -\overline{\chi_1} {X \!\!\!\!/} \chi_2\right)}  \\ 
    &-\frac{\xi}{2}(v_X+h_X)(\overline{\chi_2}\chi_2-\overline{\chi_1}\chi_1).
    \label{msat} 
\end{split}    
\end{equation}
We summarize some interesting features from the above equation. First, there is a residual $Z_2$ symmetry via Krauss-Wilczek mechanism~\cite{PhysRevLett.62.1221} {where} the $U(1)_D$ gauge symmetry is broken into its $Z_2$ subgroup~\cite{Baek:2014kna}. Only $\chi_{1,2}$ are $Z_2$-odd and can be DM candidate(s). Second, the mass splitting between $\chi_1$ and $\chi_2$ is triggered from the $\Phi^{\dag}\overline{\chi^c}\chi$ interaction after the symmetry breaking and can be written as 
\begin{equation}
    \Delta_\chi = 2\xi v_X.
\end{equation}
We then assign $M_{\chi_2} > M_{\chi_1}$ with the form, 
\begin{equation}
    M_{\chi_{1,2}} = M_{\chi}\mp\xi v_X.
\end{equation}


Before closing this section, we have to mention that the scalar sector in this model is not the focus of this work. More details for the scalar sector in fermionic inelastic DM models and relevant search strategies can be found in Ref.~\cite{Duerr:2020muu,Kang:2021oes,Li:2021rzt}. We can properly choose model parameters to satisfy all constraints from the scalar sector part in this study. 
}

\section{Signatures of inelastic DM at colliders}
\label{sec:signature}

We will discuss the production of inelastic DM at the LHC and classify the signal signatures depending on the decay length of $\chi_2$. First of all, the UFO model file of the inelastic DM model is generated by \textsf{FeynRules}~\cite{Christensen:2008py} and then we apply \textsf{MadGraph5\_aMC@NLO}~\cite{Alwall:2014hca} to generate Monte Carlo events and calculate cross sections for the following signal process, 
\begin{equation}
    pp \rightarrow A'\rightarrow \chi_2\chi_1.
\end{equation}
We consider a centre-of-mass energy of $\sqrt{s}=14$ TeV and fix the following model parameters, 
\begin{equation} 
    M_{Z'} = 3 M_{\chi_1},\quad \alpha_D = \frac{g_D^2}{4\pi} = 0.1,\quad \epsilon = 0.01,
\end{equation}
but vary $M_{\chi_1}$ and $\Delta_{\chi}$ in the range below, 
\begin{equation} 
    M_{\chi_1}/\text{GeV} = \left[ 5, 100\right],\quad \Delta_{\chi}/\text{GeV} = \left[ 0.05, 10\right], 
\end{equation} 
with the step length $5$ GeV and $0.01$ GeV for $M_{\chi_1}$ and $\Delta_{\chi}$, respectively.

\begin{figure}[t!]
\centering
\includegraphics[width=16cm,height=8cm]{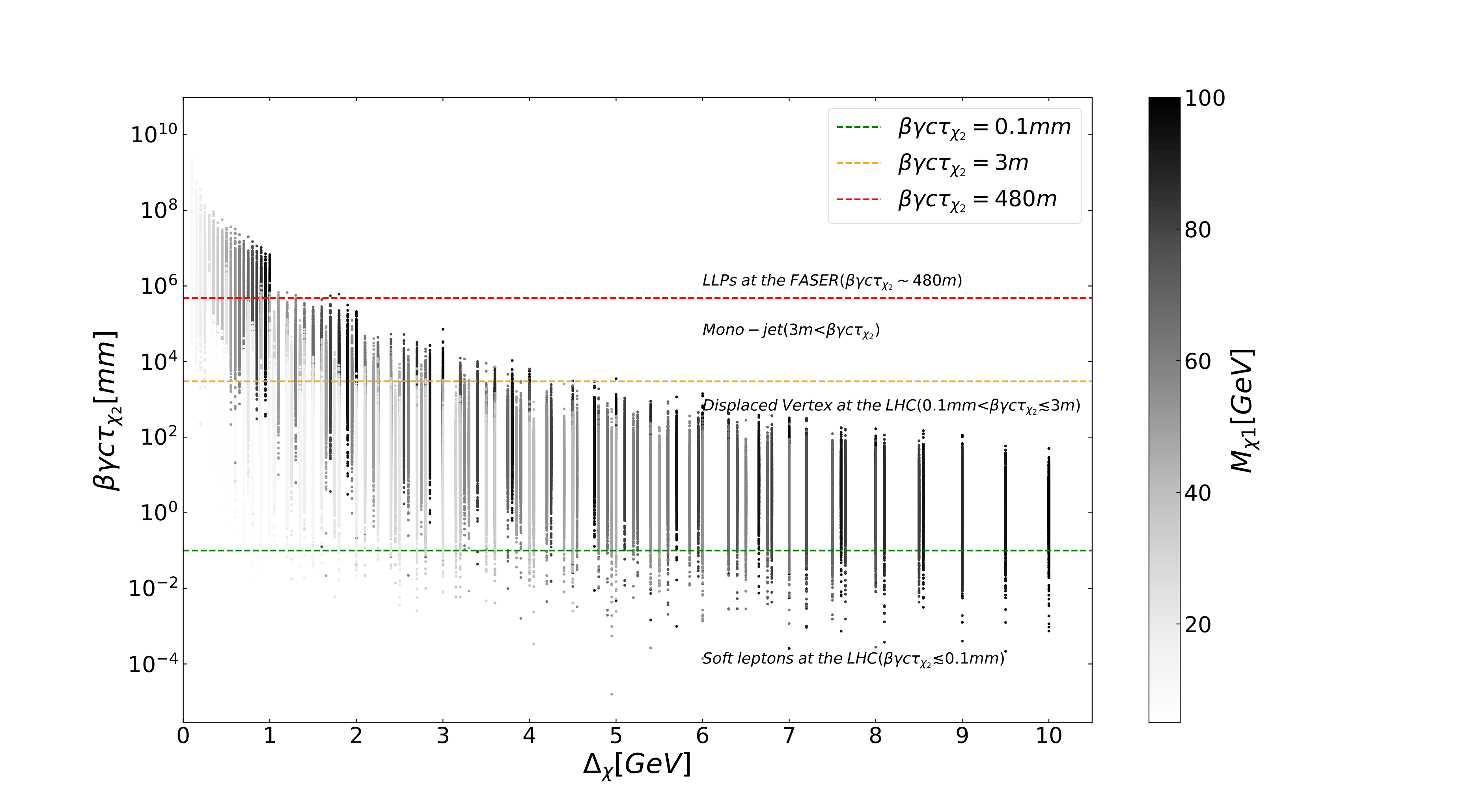}
\caption{The relationship of $(\Delta_{\chi}, \beta\gamma c\tau_{\chi_2})$ with varying $M_{\chi_1}$. The three dashed lines represent different special lab frame decay length numbers, $\beta\gamma c\tau_{\chi_2} = 0.1 \text{mm}$ (green), $\beta\gamma c\tau_{\chi_2} = 3 \text{m}$ (orange), and $\beta\gamma c\tau_{\chi_2} = 480 \text{m}$ (red). }
\label{fig:chi2_length}
\end{figure}

The time of flight for $\chi_2$ is automatically calculated in the Madgraph5@NLO. {However, when $\Delta_{\chi}\lesssim 1 \text{GeV}$, the decay widths of $\chi_2$ are adjusted using R ratio data from Particle Data Group (PDG)~\cite{Workman:2022ynf}.} In the approximation $M_{Z'}\gg M_{\chi_2}\sim M_{\chi_1}\gg M_l$, the partial decay rate for $\chi_2\rightarrow\chi_1 l^+l^-$ can be written as~\cite{Izaguirre:2015zva} 
\begin{equation} 
    \Gamma (\chi_2\rightarrow\chi_1 l^+l^-)\simeq\frac{4\epsilon^2 \alpha_{\text{em}}\alpha_D\Delta^5_{\chi}}{15\pi M_{Z'}^4} 
\label{chi2_width}    
\end{equation}
where $l = e, \mu$ and $\alpha_{\text{em}}\simeq 1/137$ is the fine structure constant. It's clear to see that once we reduce the values of $\epsilon$, $\Delta_{\chi}$, and $M_{\chi_1}$, the lifetime of $\chi_2$ will enhance.  
In the Fig.~\ref{fig:chi2_length}, we display the relations of $M_{\chi_1}$ and $\epsilon$ to the $\chi_2$ decay length. We can find the behaviors in numerical results are consistent with the approximated formula in Eq.~(\ref{chi2_width}). 

In addition, we use three dashed lines in Fig.~\ref{fig:chi2_length} to illustrate our search strategies for inelastic DM at collider experiments. Specifically, if the lab frame decay length of $\chi_2$ is as long as $\beta\gamma c\tau_{\chi_2}\simeq \mathcal{O}(500)$ m, the FASER is an ideal detector to search for inelastic DM as shown in the red dashed line in Fig.~\ref{fig:chi2_length}. Here, $\gamma$ is the Lorentz factor, $\beta$ is the velocity of $\chi_2$, and $\tau_{\chi_2} = 1/\Gamma_{\chi_2}$ is the proper decay time of $\chi_2$. 
The details for this analysis can be found in Sec.~\ref{subsec:4a}. If $\chi_2$ generates the displaced vertex at the LHC with $0.1 $ mm $ < \beta\gamma c\tau_{\chi_2}\lesssim 3$ m, the displaced muon-jet (DMJ) signature is sensitive to search for inelastic DM in this parameter space as shown in the regions below the orange dashed line in Fig.~\ref{fig:chi2_length}. We study this possibility in Sec.~\ref{subsec:4b}. Furthermore, if $\chi_2$ is prompt decay ($\beta\gamma c\tau_{\chi_2}\lesssim 0.1$ mm), the soft leptons searches at the LHC can be applied to this situation as shown in the regions below the green dashed line in Fig.~\ref{fig:chi2_length}. We recast the ATLAS analysis~\cite{ATLAS:2019lng} for soft leptons signatures in Sec.~\ref{subsec:4c}. Finally, if $\chi_2$ is the LLP with $\beta\gamma c\tau_{\chi_2} > 3$ m, the mono-jet searches can indirectly impose constraints on this model as shown in the regions above the orange dashed line in Fig.~\ref{fig:chi2_length}. However, the mono-jet constraints from the LHC are much weaker than the above ones, so we will not show the recasting for these constraints in this work. Finally, as a complementary study to cover the searches of sub-GeV inelastic DM, we utilize the mono-photon signature to search for inelastic DM at STCF. More details for this analysis will be presented in Sec.~\ref{subsec:4d}. 


\subsection{LLPs at the FASER}
\label{subsec:4a}

As we know, the B-factories can explore LLPs, but the restriction of their center-of-mass energies forces the upper bound of the mass of LLPs to be less than about $10$ GeV~\cite{Acevedo:2021wiq}. On the other hand, the ATLAS/CMS detectors at the LHC are not sensitive to the new particles with mass less than $\mathcal{O}(10)$ GeV. Therefore, a new lifetime frontier detector to search for $\mathcal{O}(10)$ GeV BSM LLPs is needed. 

In this subsection, we introduce a new detector called FASER, which has been built around the LHC to study LLPs that interact with SM particles weakly and have light masses \cite{Battaglieri:2017aum}. These particles have attracted significant attention as they could potentially explain DM and reconcile discrepancies between theoretical predictions and low-energy experiments \cite{PhysRevLett.116.042501,Feng:2017uoz,PhysRevD.73.072003,BOEHM2004219}.
Traditional detectors at the LHC primarily focus on 
heavier new particles in the central regions 
and lack the necessary sensitivity to detect light, weakly-coupled particles that are produced in the forward direction. Additionally, these particles, known as LLPs, can be highly boosted in the forward direction, traveling a macroscopic distance before decaying. Therefore, a detector located along the beamline axis in the forward region could enable the detection of light LLPs. The FASER experiment aims to address this by constructing a detector  which is 480 meters downstream from the ATLAS interaction point (IP). 
Furthermore, there is a proposal for FASER 2, which would be constructed from 2024-2026 to collect data during the HL-LHC era from 2026 to 2035 \cite{FASER:2019aik}. 
In the current investigation, we assume that LLPs produced near the IP travel along the beam axis and decay into SM particles, which can be detected by FASER. Therefore, LLPs within the acceptance angle of FASER should have high energies in the TeV range, as the decay products from LLPs would also possess energies close to the TeV scale. The full process is described as follows:
\begin{equation}
    \text{pp} \rightarrow \chi_{2} + \chi_{1},\ \chi_{2} \ \text{travels} \sim480\text{m},\ \text{then} \ \chi_{2} \rightarrow \ \chi_{1} f\overline{f}.
\end{equation}
The FASER detector is located in a region surrounded by rock, and the forward LHC infrastructure, including magnets and absorbers, helps to suppress potential background processes. Detailed simulations using FLUKA technology \cite{Ferrari:2005zk,BOHLEN2014211} have confirmed low radiation levels in LHC tunnels, with the radiative process associated with muons being the dominant background. To further mitigate backgrounds, a scintillating charged particle veto layer is employed in front of the detector \cite{Berlin:2018jbm}. Specifically, the FASER detector rejects high-energy charged particles, primarily muons, and protons to minimize additional troublesome backgrounds. With these technical measures, the background levels are considered negligible.

To enhance the trigger efficiency at low energies, FASER requires a significant deposition of visible energy from the decay products of $\chi_2$, with ${E_{\text{vis}}} >$ 100 GeV. The specific parameters for the two-phase detectors, FASER and FASER 2, are cylindrical in shape, characterized by their length (L) and radius (R):
\begin{align}
\begin{split}
       \mathbf{FASER} &: L=1.5\text{m},\ R=0.1\text{m}, \\
       \mathbf{FASER\ 2} &: L=5\text{m},\ R=1\text{m}. 
\end{split}
\end{align}
Additionally, the integrated luminosity, $\mathcal{L}$, for FASER and FASER 2 is $150\ \text{fb}^{-1}$ and $3\ \text{ab}^{-1}$, respectively.

\subsection{Displaced Muon-Jet at the LHC}
\label{subsec:4b} 

In this subsection, we focus on the signature of DMJ~\cite{Baumgart:2009tn,Cheung:2009su,Falkowski:2010cm,Izaguirre:2015pga,Kim:2016fdv,Dube:2017jgo,Zhang:2021orr} from inelastic DM models at the LHC. In the scenario where $M_{Z'} > M_{\chi_1} + M_{\chi_2}$, the $Z'$ can be on-shell produced in association with a QCD jet at the LHC~\cite{CMS:2023bay}. Subsequently, the $Z'$ decays into $\chi_1$ and $\chi_2$, and within the tracker system of the ATLAS and CMS detectors, the $\chi_2$ particle further decays into $\chi_1$ and two muons. However, due to the high boost of $\chi_2$, the resulting pair of muons from its decay becomes highly collimated, making it challenging to pass the muon isolation criteria. This phenomenon gives rise to a novel object known as a muon-jet. Our particular interest lies in a displaced dimuon vertex associated with a jet and missing momentum. This process is referred to as the DMJ signature, which is considered a particularly clean signal.

The search strategy for this kind of signature was proposed by Ref.~\cite{Izaguirre:2015zva,Berlin:2018jbm}.
Our analysis of the DMJ signature follows the methodology outlined in the above two references. In their work, most of the relevant backgrounds were found to be relatively negligible. It is worth noting that displaced vertex tracks can also be present in QCD-initiated processes, which may involve the production and subsequent decay of LLPs such as $B$ or $K$ hadrons into $\pi$ and $\mu$. The authors of Ref.~\cite{Izaguirre:2015zva} assume that the probability of such events is small but provide an approximate upper bound on the probability of a QCD-initiated event producing a hard leading jet with transverse momentum $p^j_T > 120$ GeV and two displaced muons with transverse momenta $p^{\mu}_T > 5$ GeV, considering the small mass splitting between $\chi_2$ and $\chi_1$. Additionally, in order to ensure the displacement of the muon tracks, a minimum transverse impact parameter $d_\mu > 1$ mm is imposed. Furthermore, it is known that if the decay length of $\chi_2$ is sufficiently long to allow the production of two muons will through completely the tracking system, allowing for more precise track reconstruction. Hence, a requirement is imposed that the radial displacement ($R_{\chi_2}^{xy}$) of the $\chi_2$ decay vertex is less than $30$ cm. In summary, the selection criteria for the signal region of DMJ signature encompass the following conditions: 
\begin{align}
    \begin{split}
        \mathbf{DMJ}:&\ p_T^j\ >\ 120\ \text{GeV}, \\
        &\ p_T^{\mu}\ >\ 5\ \text{GeV}, \\
        &\ d_{\mu} \ > \ 1\ \text{mm}, \\
        &\ R_{\chi_2}^{xy}\ <\ 30\ \text{cm}.
    \end{split}
\end{align}
The selection criteria outlined above have been carefully chosen in anticipation of an integrated luminosity of $\mathcal{L} = 3 \text{ ab}^{-1}$ at the High-Luminosity LHC. It is worth noting that extensive studies conducted by the authors of Ref.~\cite{Izaguirre:2015zva} have demonstrated that these criteria effectively reduce the backgrounds to a negligible level. 
\subsection{Soft Lepton Pair at the LHC}
\label{subsec:4c}
In this subsection, we investigate the search for inelastic DM models through the soft lepton pair(SLP) analysis. In our concerned process, the final state particles consist of two leptons are originated from $\chi_{2}$ decay via the off-shell $Z'$ boson. We utilize the data collected by the ATLAS detector, corresponding to an integrated luminosity of $\text{139 fb}^{-1}$ at $\sqrt{s} = \text{13}$ TeV, as described in Ref.~\cite{ATLAS:2019lng}, events with missing transverse momentum and two same-flavor, oppositely charged, low transverse momentum leptons are selected. To ensure consistency with the ATLAS analysis, we employ the analysis file provided by the \textsf{CheckMATE2} program package~\cite{Dercks:2016npn}. The \textsf{CheckMATE2} program allows us to determine whether the processes involved in inelastic DM models are excluded or not at a $95\%$ Confidence Level (C.L.), by comparing them with the results reported in Ref.~\cite{ATLAS:2019lng}. 
In our simulations, event samples are generated using \textsf{MadGraph5\_aMC@NLO}(version 2.7.2). We didn't put any pre-selections in parton level event generation, but the ME–PS matching was performed using the CKKW-L merging~\cite{Lonnblad:2011xx} scheme with the merging scale set to 15 GeV. To enforce an initial state radiation(ISR) topology, at least one parton in the final state was required to have a transverse momentum greater than 50 GeV. 
After the event reconstruction, all events entering the signal regions (SRs) undergo a common set of event selections, which is summarized in Table~\ref{table.1}.

\begin{table}[ht!] \begin{center}
 
 \begin{tabular}{l|p{8cm}} \toprule[1pt]
 
  Variable                                                                       & Event selection           \\ \hline
  Number of leptons                                                              & = 2 leptons   \\
  Leading lepton $p_T$ [GeV]                                              & $p^{\ell_1}_T > 5$       \\
  $\Delta R_{\ell \ell}$                                 &$\Delta R_{ee} > 0.3, \ \Delta R_{\mu \mu} > 0.05$      \\
  Lepton charge and flavor                                                       & $e^{\pm}e^{\mp}, \mu^{\pm}\mu^{\mp}$     \\
   $J/\psi$ invariant mass veto [GeV]                                                  & veto $3.0 < m_{\ell \ell} < 3.2$      \\
   Lepton invariant mass [GeV]                                       & $3 < m_{ee} < 60, \ 1 < m_{\mu\mu} < 60$    \\
  $E_T^\mathrm{miss}$ [GeV]       & $\geq 120$ \\
  $m_{\tau\tau}$ [GeV]                   & $m_{\tau\tau} < 0$ or $m_{\tau\tau} > 160$    \\
  Number of jets                  & $\geq 1$  \\
  Number of $b$-tagged jets                                                    & = 0     \\
  Leading jet $p_T$ [GeV]         &   $\geq 100$   \\
  min$(\Delta\phi(\mathrm{jets}, \mathbf{p}_T^\mathrm{miss}))$                          &$ > 0.4$     \\
  $\Delta\phi(j_{1}, \mathbf{p}_T^\mathrm{miss})$        & $\geq 2.0$   \\   

  \bottomrule[1pt]
 \end{tabular}\end{center}
 \caption{The event selection requirements applied to all events entering SRs for soft lepton pair analysis.}
 \label{table.1}
\end{table}
According to Table~\ref{table.1}, the event selections for our signal events require exactly two leptons of the same flavor with opposite charges. We order the leading lepton ($\ell_1$) and subleading lepton ($\ell_2$) by the size of their transverse momentum. The $p_T^{\ell_1}$ is required to be larger than $5$ GeV, which helps to reduce backgrounds from fake/nonprompt (FNP) leptons. And the subleading lepton'$p_{T}$ ($p_T^{\ell_2}$) will have different constraints on different SRs. The lepton pair is also required to have a separation $\Delta R_{\ell \ell}$, with $\Delta R_{\mu \mu}$ larger than 0.05 for a muon pair and $\Delta R_{ee}$ larger than 0.3 for an electron pair. This requirement improves the efficiency of event reconstruction by avoiding overlapping electron showers in the electromagnetic calorimeter. The final state leptons must have opposite charge and same flavor. Furthermore, the invariant mass of the lepton pair ($m_{\ell \ell}$) should fall outside the range [3.0, 3.2] GeV, which removes contributions from the $J/\psi$ decays. The $m_{\ell \ell}$ is also required to be less than 60 GeV to reduce contributions from on-shell $Z$ boson decays. Requirements on the minimum angular separation between the lepton candidates ($\Delta R_{\ell\ell}$) and $m_{\ell\ell}$ remove events in which an energetic photon produces collinear lepton pairs. The variable $m_{\tau\tau}$ represents the invariant mass approximation of a pair of $\tau$ leptons undergoing leptonically decaying processes. It is defined as $m_{\tau \tau} = \text{sign}(m_{\tau \tau}^2)\sqrt{|m_{\tau \tau}^2|}$, which is the signed square root of $m_{\tau \tau}^2 \equiv 2p_{\ell_{1}}p_{\ell_{2}}(1+\zeta_1)(1+\zeta_2)$, where $p_{\ell_{1}},p_{\ell_{2}}$ are four-momentum of two leptons, while $\zeta_1 , \zeta_2$ are the parameters in solving $\mathbf{p}_\text{T}^\text{miss} = \zeta_1 \mathbf{p}_\text{T}^{\ell_1}+\zeta_2 \mathbf{p}_\text{T}^{\ell_2}$. In certain events, the $m_{\tau\tau}$ variable can be less than zero. This occurs when one of the lepton momenta has a smaller magnitude compared to the transverse missing energy ($E_T^\text{miss}$) and points in the hemisphere opposite to the momentum imbalance vector ($\mathbf{p}^\text{miss}_\text{T}$).  
In order to mitigate backgrounds originating from the Z boson decays into $\tau^+\tau^-$, events falling within the range of $0 < m_{\tau \tau} < 160$ GeV are excluded. This selection criterion achieves an efficiency exceeding 80 percent for the analyzed signals. 

Additionally, for the events in our research process, almost invisible momentum is carried by $\chi_{1}$, these requirements on $E_T^\mathrm{miss}$ suggest that the process our concerned is recoiling against additional hadronic activities, like ISR. All events are therefore required to have at least one jet with $p^j_{T}>100$ GeV, therefore, the missing transverse energy ($E_T^\mathrm{miss}$) is required to be greater than $120$ GeV in this analysis, even higher in some SRs. Additional jets in the event are also required to be separated from the $\mathbf{p}_{T}^\mathrm{miss}$ by $\text{min}(\Delta\phi(\text{jets}, \mathbf{p}_T^\mathrm{miss}))>0.4$ in order to suppress the impact of jet energy mismeasurement on $E_T^\mathrm{miss}$. 
Our research process, focusing on the final state particles with two leptons, events with one or more $b$-tagged jets are vetoed to reduce backgrounds from SM $t\bar{t}$ production. 

\subsection{Mono-photon Event at the STCF}
\label{subsec:4d}
In this subsection, we discuss the search for the light $Z'$ decay to $\chi_1\chi_2$ via the mono-photon signature at the future STCF which is an $e^+e^-$ collider project with a peak luminosity of $10^{35}$ $\text{cm}^{-2}\text{s}^{-1}$ and operating in $\sqrt{s} = 2$ to $7$ GeV~\cite{Barniakov:2019zhx,Liang:2021kgw}. The process of interest is $e^{+} e^{-} \rightarrow \gamma Z^{\prime} \rightarrow \gamma (\chi_1 \chi_2)$. However, this process is subject to both reducible and irreducible backgrounds due to the limited detection capability. The main reducible SM backgrounds include the processes $e^{+}e^{-} \rightarrow \gamma f \overline{f}$ and $e^{+}e^{-} \rightarrow \gamma \gamma (\gamma)$, where the final state particles are emitted in the solid angle region not covered by the detectors. 
The process $e^{+}e^{-} \rightarrow \gamma e^{+}e^{-}$, where the final state electron and positron are collinear with the beam directions, receives a significant contribution from $t$-channel diagrams. The irreducible SM backgrounds to our process are the processes $e^{+}e^{-} \rightarrow \gamma \nu_{\ell} \overline{\nu}_{\ell}$, where $\nu_{\ell} = \nu_{e}, \nu_{\mu},\nu_{\tau}$ are SM neutrinos. 
In this study, we apply specific cuts on the final state photon to reduce background events. These cuts are based on the analysis from BESIII~\cite{PhysRevD.96.112008} and are used for both reducible and irreducible SM backgrounds. Specifically, we impose the conditions $E_{\gamma} > 25$ MeV in the barrel region ($|z_{\gamma}| < 0.8$) and $E_{\gamma} > 50$ MeV in the end-caps region ($0.92 > |z_{\gamma}| > 0.86$), where $E_{\gamma}$ is the photon energy and $z_{\gamma} \equiv \cos{\theta_{\gamma}}$ with $\theta_{\gamma}$ being the relative angle between the electron beam axis and the photon momentum. However, applying these cuts alone does not effectively suppress the contribution from reducible backgrounds, which remains significant. To further address this, we introduce an additional cut based on momentum conservation in the transverse direction and energy conservation~\cite{PhysRevD.100.115016, Liu:2019ogn,Zhang:2019wnz}. As an example, considering the reducible background process $e^{+}e^{-} \rightarrow \gamma e^{+}e^{-}$, we use energy conservation in the center-of-mass frame and transverse momentum conservation to obtain the following relations:
\begin{equation}
     E_{\gamma} + E_{1} + E_{2} = \sqrt{s}, 
\end{equation}       
\begin{equation}
E_{\gamma}\sin{\theta_{\gamma}} - E_{1}\sin{\theta_{1}} - E_{2}\sin{\theta_{2}}= 0.
\end{equation}
Here, $E_{\gamma}$, $E_{1}$, and $E_{2}$ are energies of the final state photon, electron, and positron, respectively, and $\theta_{\gamma}$, $\theta_{1}$, and $\theta_{2}$ represent their respective polar angles. By imposing the condition that both $|\cos{\theta_{1,2}}|$ are greater than or equal to $|\cos{\theta_{b}}|$, where $|\cos{\theta_{b}}|$ is the polar angle at the boundary of the sub-detector where the final state electron and positron are emitted. We request $|\cos{\theta_{b}}| \geq 0.95$, and the final photon energy cut,
\begin{equation}
    E_{\gamma} \ > \ E_{b}(\theta_{\gamma}) \ = \ \frac{\sqrt{s}}{(1+\frac{\sin{\theta_{\gamma}}}{\sin{\theta_{b}}
    })}. 
\end{equation}
The energy cut $E_{b}$ is determined by the polar angle $\theta_{b}$ and ensures that the final state photon lies outside the boundary region. In this work, we aim to probe the light $Z'$ decay to $\chi_1\chi_2$ via the mono-photon signature at $\sqrt{s} = 4$, $7$ GeV, 
and corresponding to $\mathcal{L} = 30~\text{ab}^{-1}$.

\subsection{Numerical Results and Discussions} 
\label{sec:result}

\begin{figure}[t!]
\centering
\includegraphics[width=0.45\textwidth]{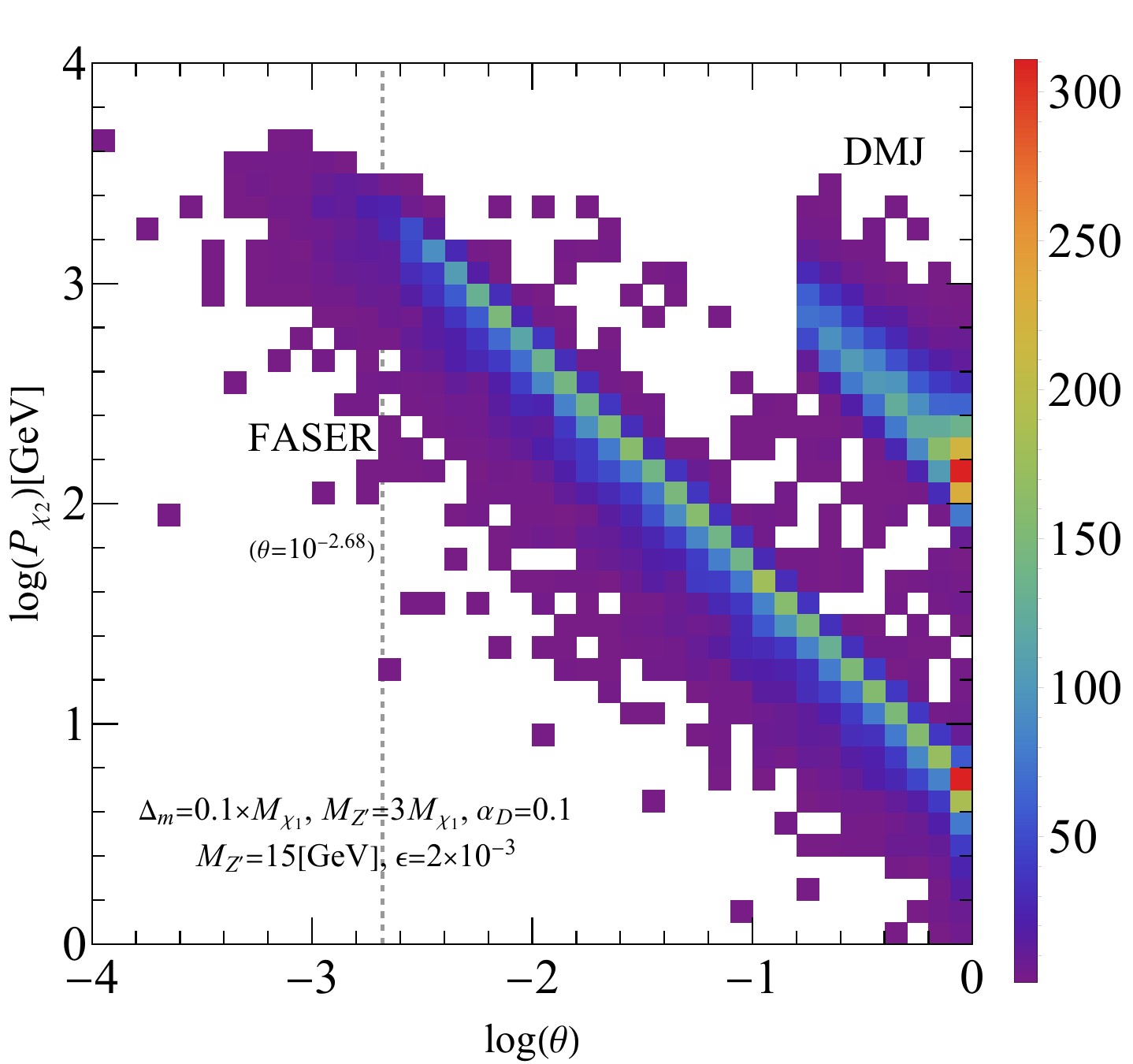}
\includegraphics[width=0.45\textwidth]{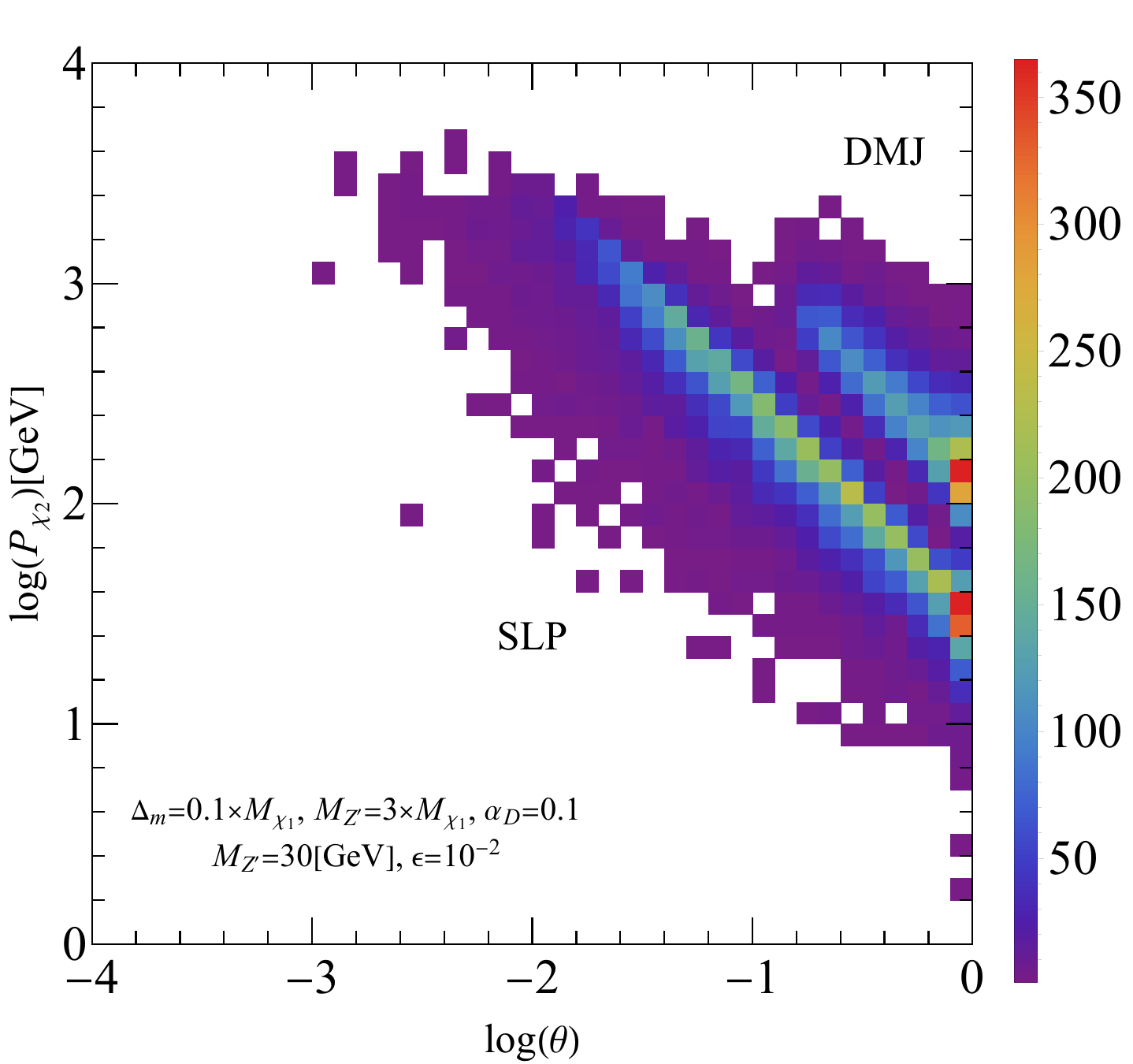}
\caption{The $P_{\chi_2}-\theta$ distribution of $\chi_2$. The left panel is under the search strategies of FASER and LHC(DMJ) with $M_{\chi_2}$ = 5.5\ GeV and $M_{Z'} = 15$ GeV. The right panel is under the search strategies of DMJ and SLP at the LHC with $M_{\chi_2}$ = 11\ GeV and $M_{Z'} = 30$ GeV. The vertical line (dashed grey) in the left panel indicates the decay length in the lab frame for FASER 2 (480 m), which tells us there exist sensitive regions with $\chi_2$ when $\theta<10^{-2.68}$.} 
\label{fig:3}
\end{figure}

In Fig~\ref{fig:3}, we present the event distribution over the magnitude of the spatial momentum $P$ of $\chi_2$ and its angle $\theta$ along the beam line under different search strategies on various benchmark points. In the left panel, the $P_{\chi_2}-\theta$ distribution in the lower left area corresponds to the LLPs search at the FASER, while the distribution in the upper right corner represents the DMJ search at the LHC. These distributions are obtained for fixed $M_{Z'} = 15$ GeV and $\epsilon = 2 \times 10^{-3}$. The dashed grey line denotes the acceptance angle for FASER 2 ($\theta < 10^{-2.68}$), illustrating that the FASER detector has sensitivity primarily to highly boosted particles with a very small angle relative to the beam line. The reason for this is that the $\chi_2$ produced in this process has a very low transverse momentum ($P_T$), which is nearly equal to $M_{\chi_2}$, and it is emitted in the forward direction, with its trajectories collimated along the beam-line. The DMJ event distribution tells us the high off-beam-line sensitivity because there exists a high $P_T$ jet($>$120 GeV) back to $Z'$. Since we only require $P_T^{\mu} > 5 \text{GeV}$ in this analysis, it leads to the lower threshold of $P_{T}^{\chi_2}$ $\geq$ 50 GeV ($P_T^{\mu} \sim \Delta \times P_{T}^{\chi_2}$). We can clearly find that the distribution pattern in the left panel indicates a kinematic overlap between these two strategies. Next, the right panel in Fig.~\ref{fig:3} corresponds to the SLP searches at 13 TeV LHC and the DMJ searches at the LHC. These distributions are obtained for fixed $M_{Z'} = 30$ GeV and $\epsilon = 10^{-2}$. The lower left $P_{\chi_2}-\theta$ distribution represents the SLP search, while the upper right part corresponds to the DMJ search. 
 We observe a larger kinematic overlap between these two strategies compared to the left panel, which is consistent with the results presented in Fig.~\ref{fig:2}.

\begin{figure}[htbp]
\centering
\includegraphics[width=0.45\textwidth]{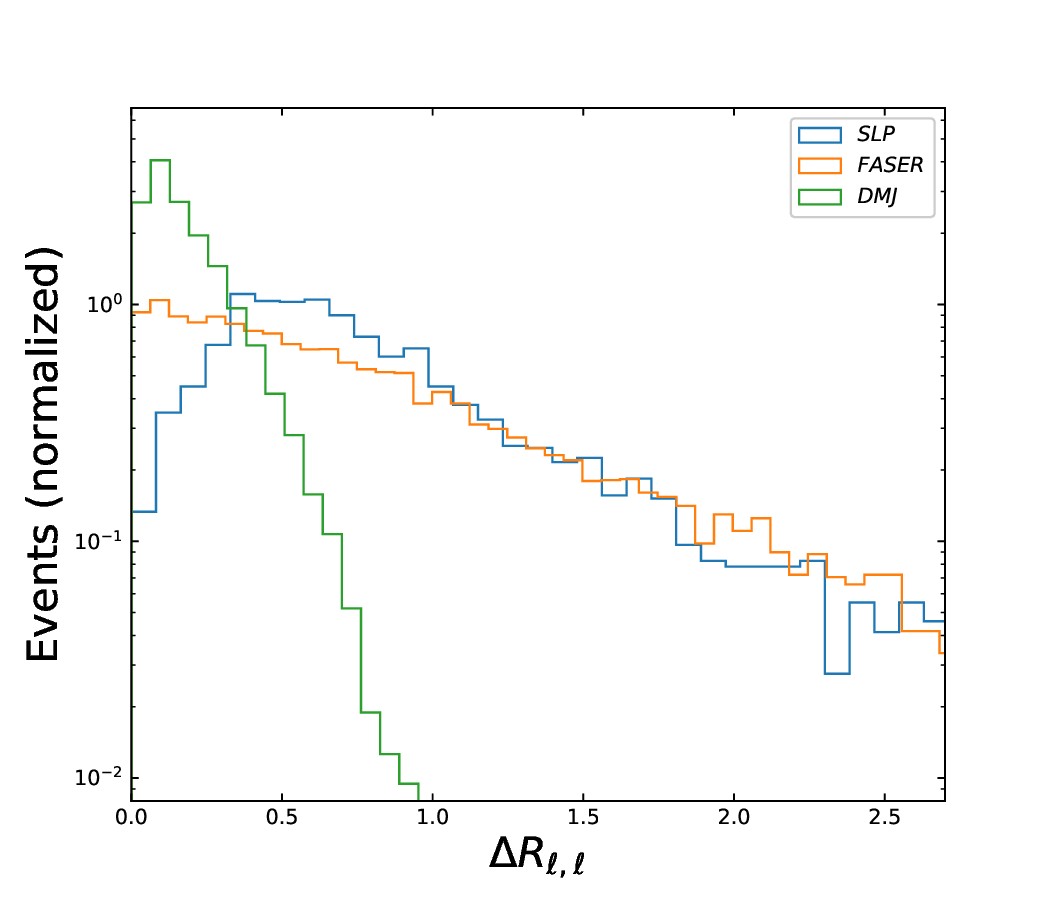}
\includegraphics[width=0.45\textwidth]{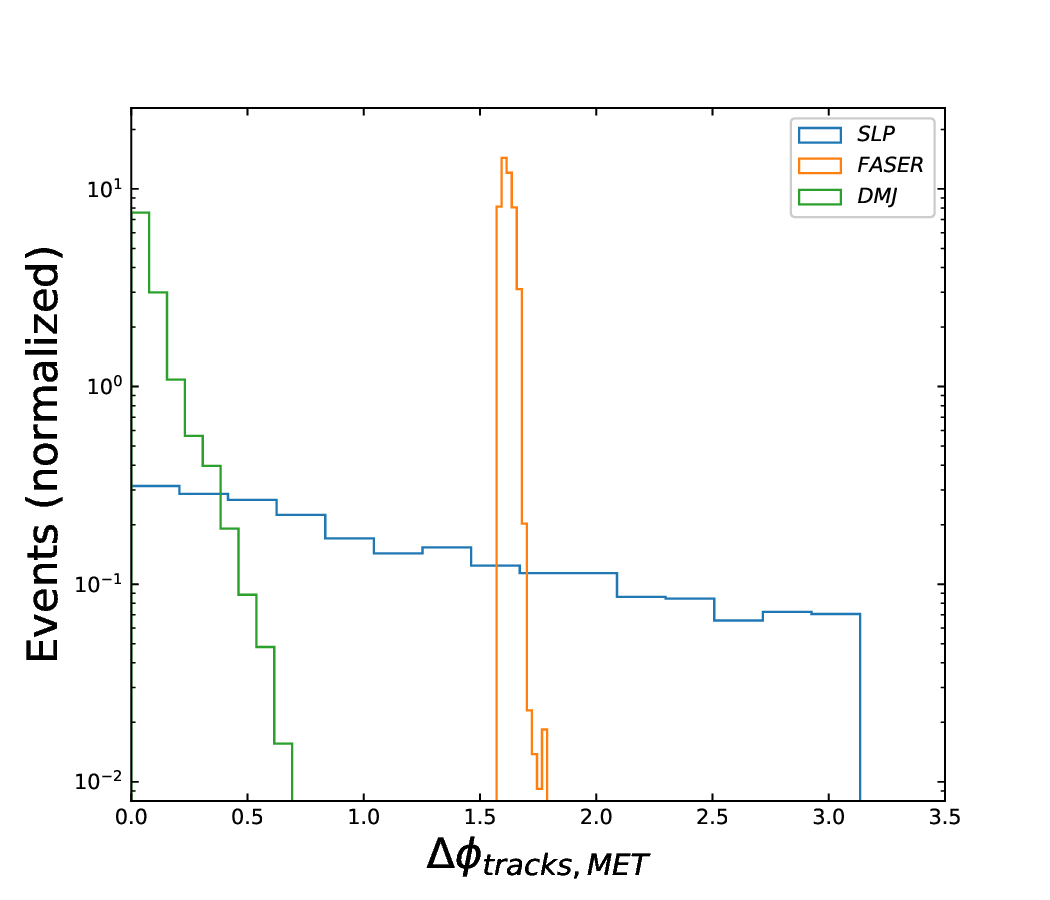}
\caption{
The left panel represents the $\Delta R$ distributions between two leptons in the final state from the $\chi_2$ decays, and the right panel shows the $\Delta \phi$ distributions between the spatial momentum of the lepton tracks and the missing energy. }
\label{fig:4}
\end{figure}

In Fig~\ref{fig:4}, we provide the kinematic distributions of $\Delta R_{\ell \ell}$ and $\Delta \phi_{\text{Tracks, MET}}$, with the fixed parameter points in their own probing regions, $M_{\chi_1} = 1$ GeV and $\epsilon = 10^{-2}$ for the FASER, $M_{\chi_1} = 10$ GeV and $\epsilon = 10^{-2}$ for the LHC (DMJ), and $M_{\chi_1} = 160$ GeV and $\epsilon = 7\times10^{-2}$ for the LHC (SLP) respectively, to illustrate their distinct kinematic properties in these three search strategies. In the left panel of Fig~\ref{fig:4}, we present the opening angle distribution of two leptons, $\Delta R_{\ell \ell}$, originating from the decay of $\chi_2$. The STCF signature is not considered in this analysis as its mass range is predominantly below the GeV scale. The DMJ search at LHC exhibits the largest collinear feature of a lepton pair compared to the other two search strategies because of the initial high $P_T$ jet in this signature. In contrast, the SLP searches at the LHC show a relatively weak collimation among the three strategies. 
The right panel of Fig~\ref{fig:4} displays the relative azimuthal angle between the lepton tracks and the missing energy, $\Delta \phi_{\text{Tracks, MET}}$. We observe that the DMJ signature at the LHC exhibits a smaller $\Delta \phi_{\text{Tracks, MET}}$ compared to the signature at the FASER. This difference arises from the fact that the DMJ process involves an initial high $p_T$ jet causing a high energy $Z'$ recoiling this jet, therefore, leading the productions of $Z'$ to be relatively collinear. In contrast, the situation at the FASER results in larger relative angles between the produced $\chi_1$ and $\chi_2$. 
For the SLP searches, which involve matching and merging effects of QCD jets, the distribution of $\phi_{\text{Tracks, MET}}$ tends to be more evenly distributed. Therefore, $\Delta R_{\ell \ell}$ and $\Delta \phi_{\text{Tracks, MET}}$ distributions provide insights into the kinematic characteristics of different search strategies. 

In our analyses, we take the benchmark parameters as~\cite{Kang:2021oes,Berlin:2018jbm,Izaguirre:2017bqb,Izaguirre:2015zva,Bertuzzo:2022ozu}: $M_{Z'} = 3 M_{\chi_{1}}$, $\alpha_{D} \equiv g_{D}^{2}/4\pi =0.1$, and $\Delta_{\chi}= 0.1 M_{\chi_{1}}$ . After applying the above event selections and search strategies, we present the final results regarding the projected sensitivity for different search strategies. 
Compared with the work of~\cite{Berlin:2018jbm}, our results will be presented in the $\left(M_{\chi_1}, \epsilon\right)$ plane, covering the range $10^{-3}$ GeV $< M_{\chi_1} < 10^{3}$ GeV. This range is particularly intriguing as it includes extensive areas that are still unconstrained by current experimental measurements. It's noteworthy that their work predominantly addresses DM masses at the GeV scale and smaller mass splittings ($\Delta_{\chi}<0.1M_{\chi_1}$) within the context of various LHC lifetime frontier experiments. In contrast, our study delves into the mono-photon signature at the new low-energy $e^+e^-$ collider, STCF, extending the inelastic DM mass range into the MeV scale. We employ a soft lepton pair analysis for inelastic DM models, adapting the corresponding ATLAS experiment analysis. Our results contribute to covering a certain parameter space that has not been excluded by previous work. Additionally, we explore a broader mass splitting, $\Delta_{\chi}/M_{\chi_1}$, ranging from $0.01$ to $0.4$, corresponding to the DM co-annihilation mechanism in the early universe. Thus, our approach investigates whether DM co-annihilation with various $\Delta_{\chi}/M_{\chi_1}$ in the early universe can be tested in current and future experiments.
 

\begin{figure}[t!]
\centering
\includegraphics[scale=0.5]{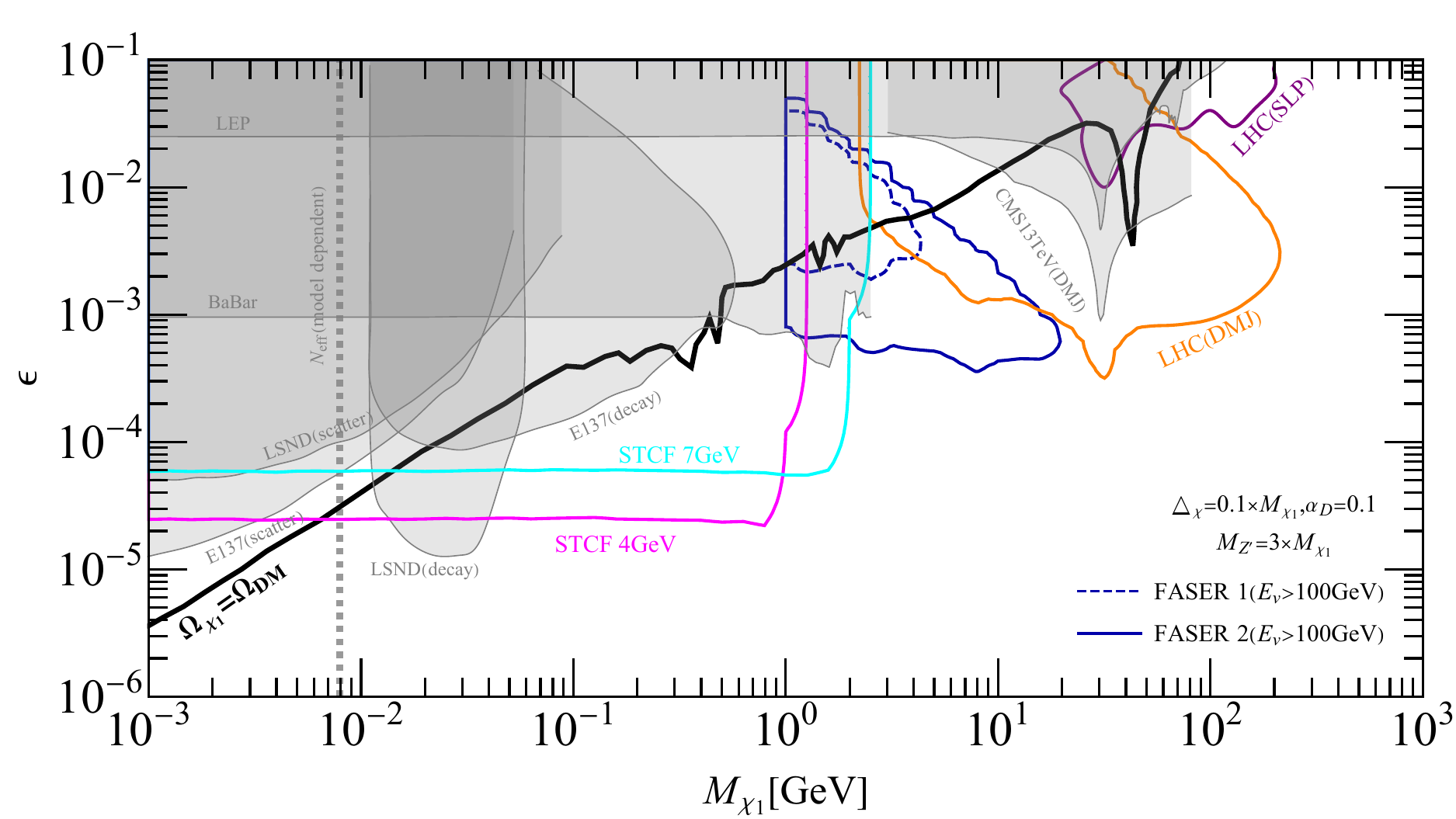}
\caption{The existing bounds (gray bulks) and sensitivities (color lines) for the fermionic inelastic DM models in the $\left(M_{\chi_1}, \epsilon\right)$ plane with fixed $\alpha_D = 0.1$, $M_{Z'}=3 M_{\chi_1}$, $\Delta_{\chi} = 0.1 M_{\chi_1}$. The black contour represents the region where the abundance of $\chi_1$ matches the observed DM relic density~\cite{Izaguirre:2015zva,PhysRevD.96.055007,Duerr:2019dmv}. The light gray regions represent the excluded regions from LEP~\cite{Hook:2010tw, Curtin:2014cca} and BaBar~\cite{BaBar:2017tiz} as well as LSND and SLAC E137 beam-dumps~\cite{Izaguirre:2017bqb}. The colored contours indicate the projected reach of different strategies. Specifically, the orange contours correspond to the reach of searches at ATLAS and CMS~\cite{PhysRevD.93.063523, Berlin:2018jbm}, while the dark dashed blue and dark blue contours represent the reach of FASER1 and FASER2~\cite{FASER:2018bac, FASER:2019aik, Feng:2017uoz}. The purple region shows the excluded region from the soft lepton pair search at ATLAS based on recast experimental analyses~\cite{ATLAS:2019lng}. The sensitivity of the STCF search via monophoton is displayed in magenta and cyan for $\sqrt{s} = 4$, $7$ GeV~\cite{Epifanov:2020elk}. 
}
\label{fig:2}
\end{figure}

Our main findings are illustrated in Fig.~\ref{fig:2}, which includes both light gray regions and colored contour regions. The light gray regions represent the current experimental constraints 
derived from experiments such as BaBar~\cite{BaBar:2017tiz}, LEP~\cite{Hook:2010tw, Curtin:2014cca}, and CMS $13$ TeV DMJ~\cite{CMS:2023bay}. Besides, the boundary lines derived from fixed target experiments E137 and LSND can be found in Ref.~\cite{Izaguirre:2017bqb}. In contrast, the colored contour regions depict the projected sensitivities obtained through the four strategies employed in our analysis\footnote{The FASER results from our analyses are slightly different from the corresponding results in Ref.~\cite{Berlin:2018jbm}. The main reasons for these differences stem from the absence of consideration for the production of DM states from meson decays, the distinction in $c_W$ resulting from the $Z'$ coupling to fermions in the SM, and the selection of different size parameters by the detectors. Here the dominant contributions to the final cross-section comes from the Drell-Yan type process, particularly when the mass of $\chi_1$ exceeds 1 \text{GeV}.}. Finally, the black bold line represents the parameter space where the abundance of $\chi_1$ agrees with the measured DM relic density~\cite{Izaguirre:2015zva,PhysRevD.96.055007,Duerr:2019dmv}. As illustrated in Fig~\ref{fig:2}, our analysis demonstrates that all the search strategies employed in this work are capable of probing the parameter space that is not excluded by current constraints.

In this study, we have chosen a fixed ratio of the $Z'$ mass to the $\chi_1$ mass, specifically $M_{Z'}/M_{\chi_1} = 3$. However, if the value of $M_{Z'}/M_{\chi_1}$ is increased, both the relic density curve and projected sensitivities line in our search strategies would shift upward compared to that in Fig.~\ref{fig:2}.  
Inversely, the relic density curve and projected sensitivities line shift downward when $M_{Z'}/M_{\chi_1}$ decreases. 

\begin{figure}[h!]
  \centering

  \begin{minipage}{0.5\textwidth}
    \includegraphics[width=\textwidth]{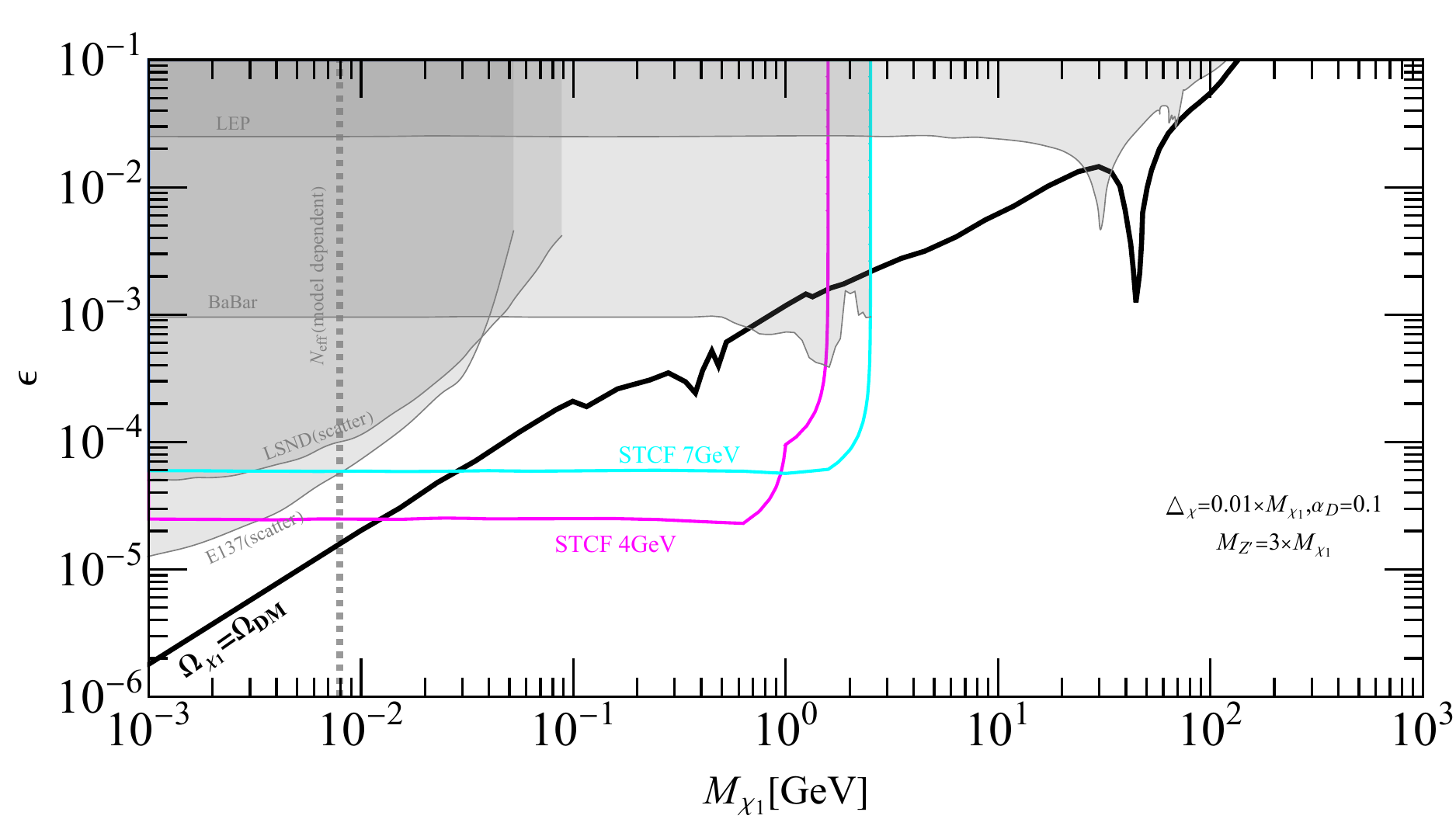}
  \end{minipage}\hfill
  \begin{minipage}{0.5\textwidth}
    \includegraphics[width=\textwidth]{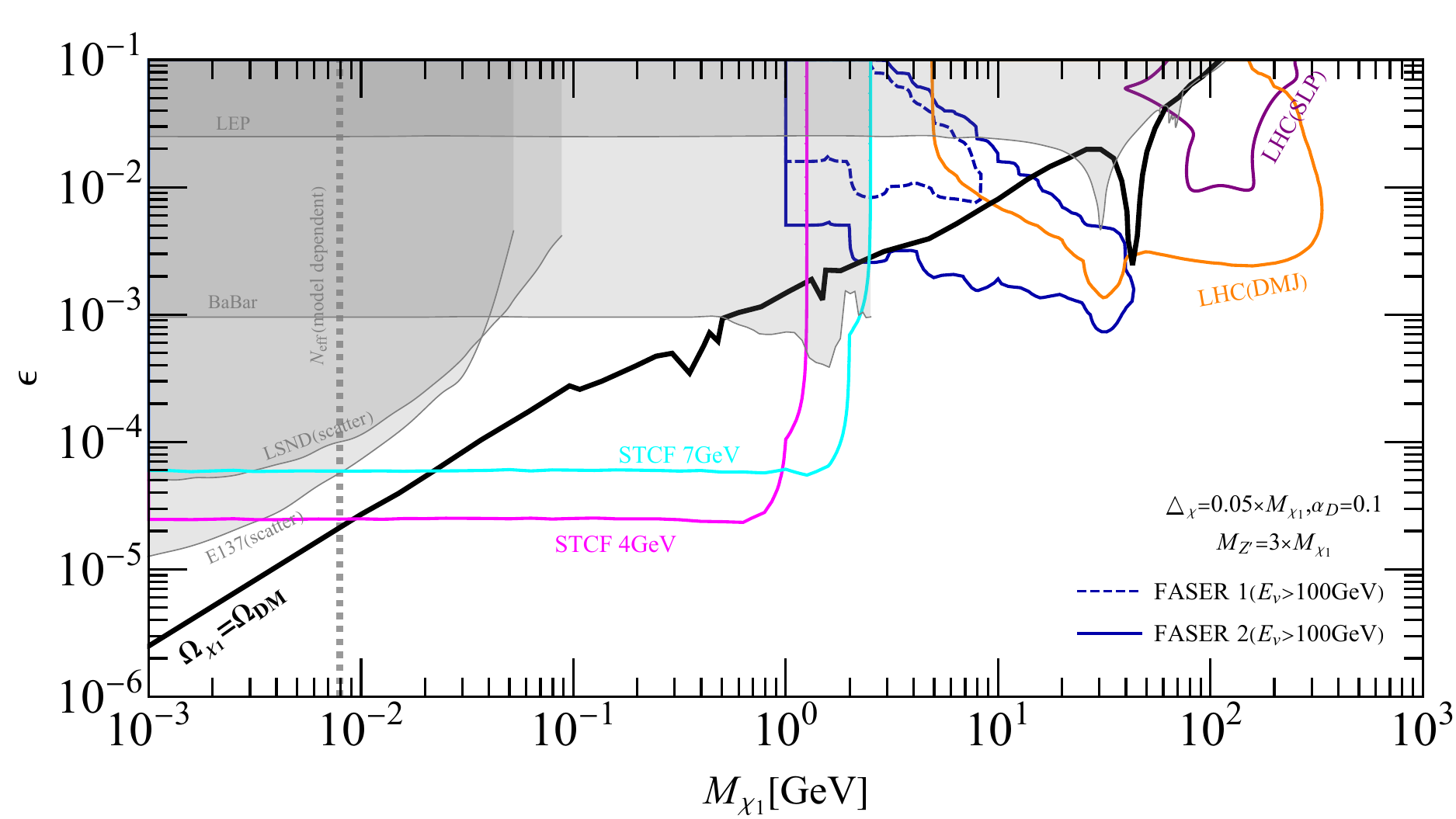}
  \end{minipage}

  \vspace{0.2cm} 

  \begin{minipage}{0.5\textwidth}
    \includegraphics[width=\textwidth]{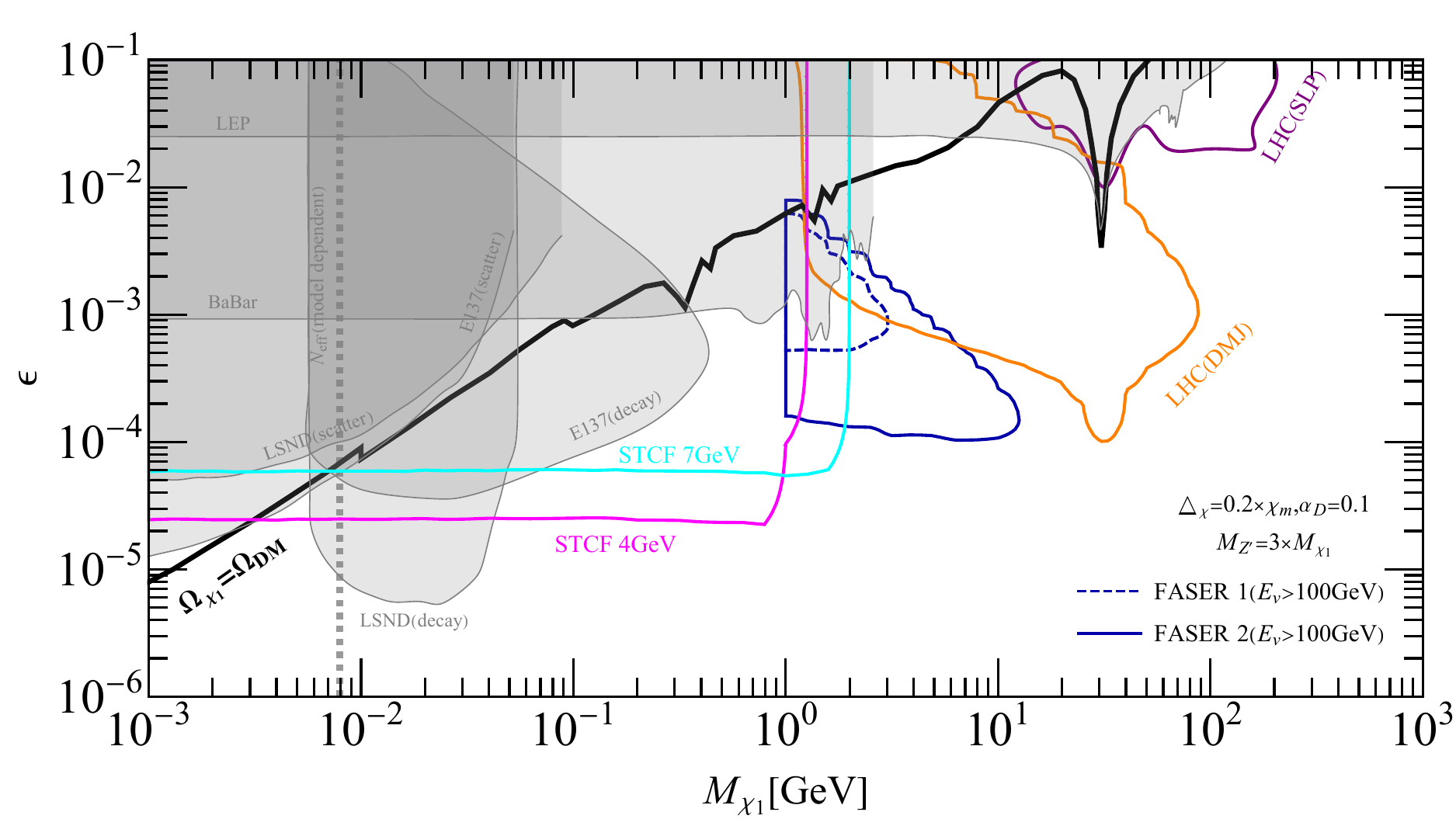}
  \end{minipage}\hfill
  \begin{minipage}{0.5\textwidth}
    \includegraphics[width=\textwidth]{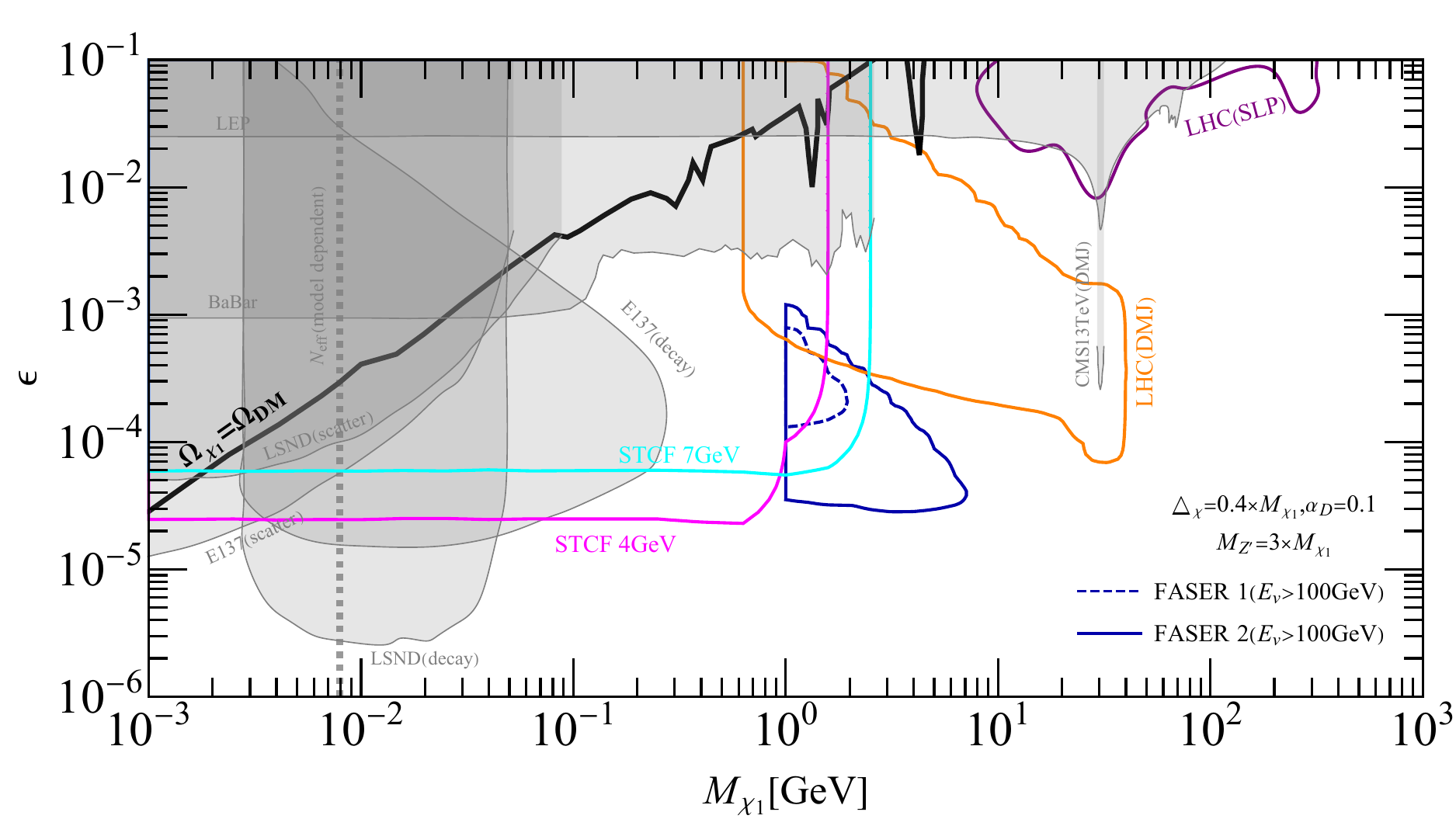}
  \end{minipage}
  \caption{Similar to Fig.~\ref{fig:2}, there are existing bounds (gray bulks) and sensitivities (color lines) for the fermionic inelastic DM models in the $\left(M_{\chi_1}, \epsilon\right)$ plane with fixed $\alpha_D$=0.1, $M_{Z'}=3 M_{\chi_1}$, but under different choices of mass splitting, $\Delta_{\chi}/M_{\chi_1} = 0.01$ (upper-left), $0.05$ (upper-right), $0.2$ (lower-left), $0.4$(lower-right). The black contour represents the region where the abundance of $\chi_1$ matches the observed DM relic density~\cite{Izaguirre:2015zva,PhysRevD.96.055007,Duerr:2019dmv}. }
  \label{fig5}
\end{figure}

On the other hand, in our initial analysis, we set the mass splitting as $\Delta_{\chi}/M_{\chi_1} = 0.1$. However, to compare the results under other mass splitting settings with our initial findings, we consistently present the final results for $\Delta_{\chi}/M_{\chi_1} = 0.01, 0.05, 0.2, 0.4$, as shown in Fig.~\ref{fig5}. It can be observed that as $\Delta_{\chi}/M_{\chi_1}$ decreases, the relic density curve shifts downward in Fig.~\ref{fig5}. Simultaneously, the projected results of different search strategies shift upward, diminishing sensitivity when $\Delta_{\chi}/M_{\chi_1}$ is small enough. 
For the case of a mass splitting $\Delta_{\chi} = 0.01 M_{\chi_1}$, as depicted in the upper-left panel of Fig.~\ref{fig5}, we observed that, owing to such a small mass splitting, the visible components from the $\chi_2$ decay are too soft to satisfy the detection requirements. Consequently, apart from the mono-photon searches at the STCF, the other three search strategies fail to yield any prospective bounds. 
Hence, exploring additional search strategies is necessary to further investigate this parameter space, such as the mono-jet search at the LHC~\cite{ATLAS:2021kxv, CMS:2021far}. Particularly, employing a larger mass splitting for SLP searches at the LHC could help exclude a larger parameter space.
 

It is important to note that while our work demonstrates good complementary among the four strategies in probing inelastic DM, there are still other exploration methods that can be pursued including time-delayed tracks at the LHC~\cite{Liu:2018wte, Berlin:2018jbm} and other LLPs experiments like MATHUSLA~\cite{MATHUSLA:2022sze}, CODEX-b~\cite{Aielli:2022awh}, AL3X~\cite{Dercks:2018wum}, and so on. 

\section{Conclusion} 
\label{sec:conclusion}

In this work, we study the prospects of searching for the inelastic DM at colliders. Due to the constraint of DM relic density, the mass splitting between the heavier and lighter dark states should be small to achieve the coannihialtion, which leads to some unconventional signatures in this model. For the inelastic DM mass in the range of 1 MeV to 210 GeV, we find that most of the parameter space that can provide the correct relic density could be probed by searching for the long-lived particles at the FASER, the displaced muon-jets and soft leptons at the LHC, and the mono-photon events at the STCF.

\section*{Acknowledgments} 

We thank Wei Liu, Yu Zhang and Bin Zhu for the helpful discussions. 
The work is supported by the National Natural Science Foundation of China (NNSFC) under grant No. 12147228 and Project 12047503 supported by NSFC.

\bibliography{refs}

 \end{document}